\documentclass[sigconf]{acmart}
\usepackage{caption}
\usepackage{subcaption}
\usepackage{graphicx}
\usepackage{hyperref} 

\AtBeginDocument{%
  \providecommand\BibTeX{{%
    \normalfont B\kern-0.5em{\scshape i\kern-0.25em b}\kern-0.8em\TeX}}}

\acmConference[SenSys 2020]{SenSys 2020: 18th ACM Conference on Embedded Networked Sensor Systems}{November 16-19, 2020}{Yokohama, Japan}

\begin{document}

\title[Tracking Candidate Therapeutics for COVID-19 in CORD-19]{Tracking Short-Term Temporal Linguistic Dynamics to Characterize Candidate Therapeutics for COVID-19 in the CORD-19 Corpus}

\author{James Powell}
\email{jepowell@lanl.gov}
\orcid{0000-0002-3517-7485}
\affiliation{%
  \institution{Los Alamos National Laboratory}
  \streetaddress{P.O. Box 1663}
  \city{Los Alamos}
  \state{New Mexico}
  \country{USA}
  \postcode{87545}
}
\author{Kari Sentz}
\email{ksentz@lanl.gov}
\affiliation{%
  \institution{Los Alamos National Laboratory}
  \streetaddress{P.O. Box 1663}
  \city{Los Alamos}
  \state{New Mexico}
  \country{USA}
  \postcode{87545}
}

\begin{abstract}
Scientific literature tends to grow as a function of funding and interest in a given field. Mining such literature can reveal trends that may not be immediately apparent. The CORD-19 corpus represents a growing corpus of scientific literature associated with COVID-19. We examined the intersection of a set of candidate therapeutics identified in a drug-repurposing study with temporal instances of the CORD-19 corpus to determine if it was possible to find and measure changes associated with them over time. We propose that the techniques we used could form the basis of a tool to pre-screen new candidate therapeutics early in the research process. 
\end{abstract}

\maketitle 

\section{Introduction}

Diachronic word analysis is Natural Language Processing (NLP) technique for characterizing the evolution of words over time. Often used for historical linguistic studies, it can also be applied to scientific literature \cite{Tshitoyan19} and can reveal early evidence of scientific discoveries before they become widely known. 

Drug-repurposing studies aim to identify existing drugs that might be useful in treating other diseases. The availability of large amounts of data about drugs and infectious agents such as viruses has enabled such studies to be performed in-silico. In early 2020, a number of repurposing studies were undertaken to identify potential treatments for COVID-19. 

\begin{figure}[H]
  \centering
  \includegraphics[width=0.4\textwidth]{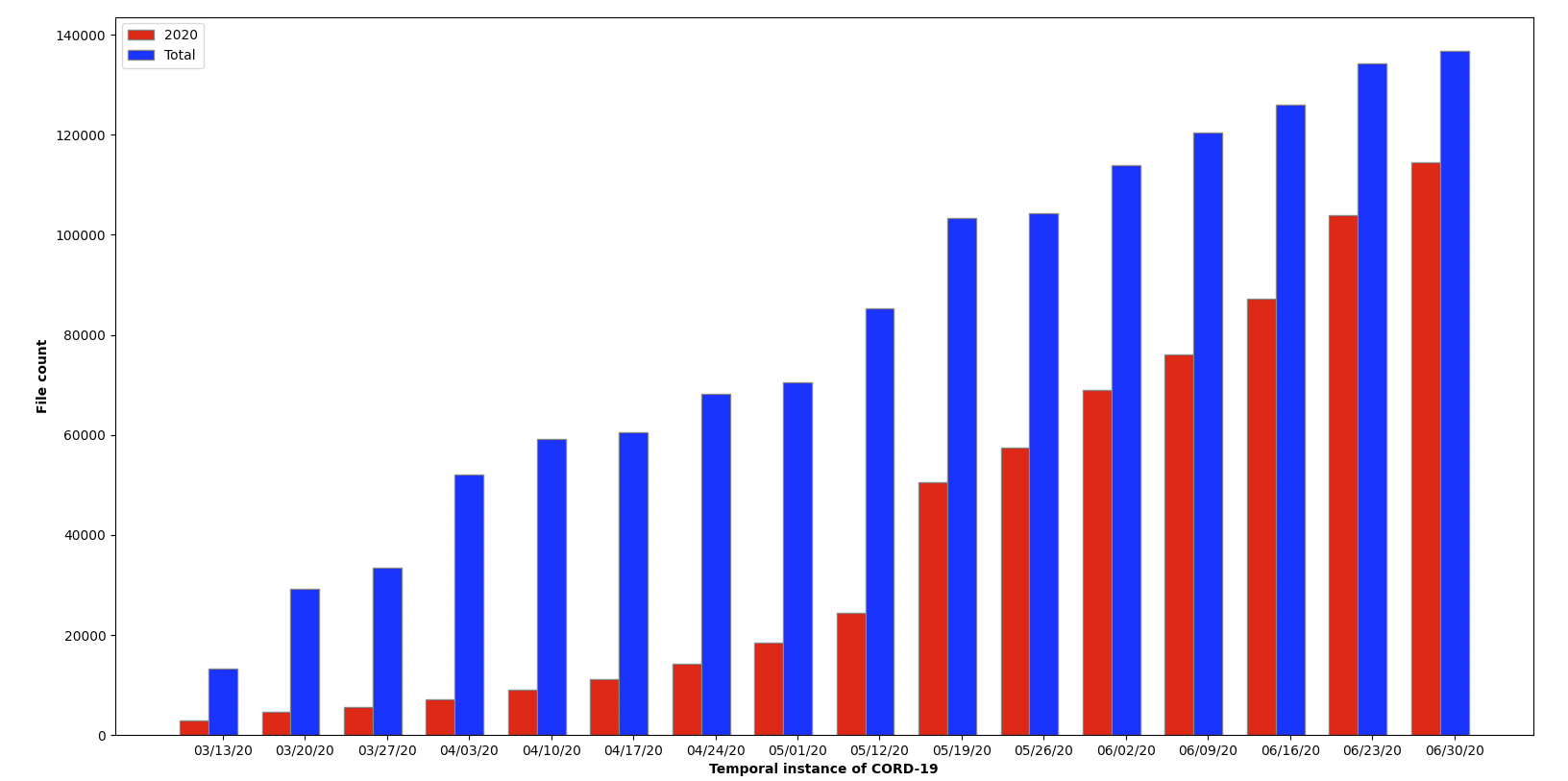}
  \caption{Weekly growth of the CORD-19 corpus}
  \label{fig:cord19}
  \Description{}
\end{figure}

The CORD-19 corpus \cite{Wang20} was established in March 2020 as a repository for research related to SARS-COV-2 and other coronaviruses. It aggregates content from PubMed, bioRxiv, medRxiv, and other sources, and it is updated with new publications on a regular basis. Figure ~\ref{fig:cord19} illustrates the growth of CORD-19 through mid-2020. 

Using CORD-19, we conducted a diachronic survey of candidate therapeutics identified in one of the more exhaustive drug re-purposing studies conducted to date for COVID-19, undertaken in February 2020 at Oak Ridge National Laboratory. This study, detailed in \cite{Smith20}, analyzed drugs in the SWEETLEAD database for potential antiviral properties. The study produced a dataset identifying over 9,000 existing approved drugs and supplements as potential candidate therapeutics for COVID-19. Most importantly for our purposes, this dataset included commonly used drug or supplement names for each candidate. 

Our survey considered the following questions:

\begin{itemize}
\item{How many candidate therapeutics appear in CORD-19?}
\item{Do references to the candidates change over time?}
\end{itemize}

\section{Materials and Methods}
Using temporal snapshots of CORD-19 spanning March 13 to June 30, we computed frequency and semantic representations for each candidate therapeutic found in the corpus. For each temporal instance, we computed TF/IDF (Term Frequency/Inverse Document Frequency) score for each candidate, a common metric to evaluate the relative importance of terms in a corpus. 

\begin{figure}[H]
  \centering
  \includegraphics[width=0.4\textwidth]{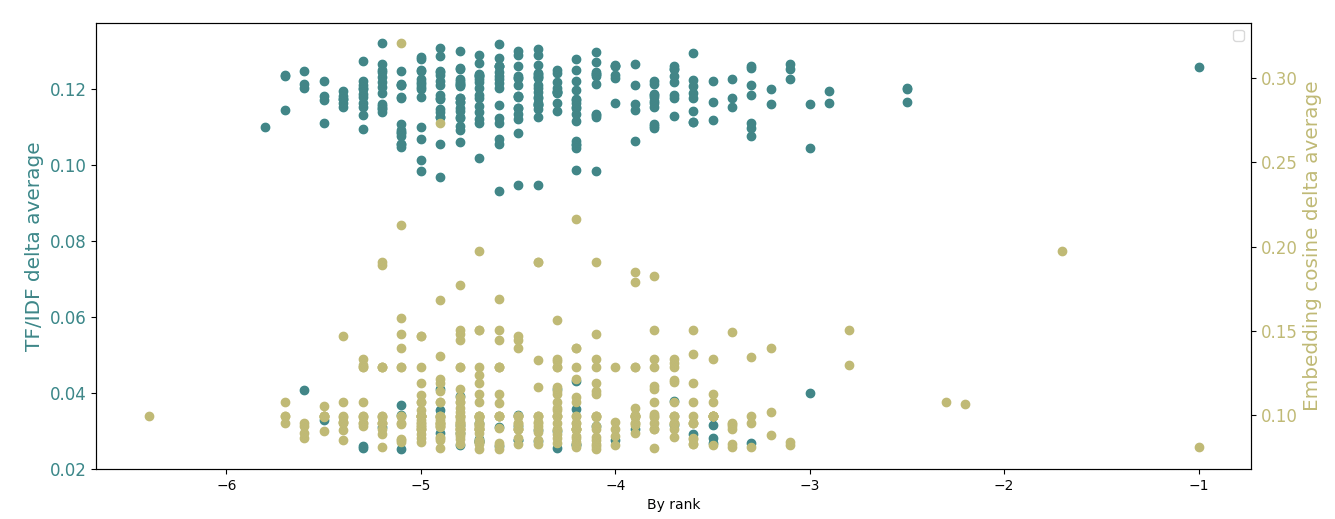}
  \caption{TF/IDF and cosine embedding distances }
  \label{fig:rankscore}
  \Description{}
\end{figure}

To perform semantic analysis, we first computed diachronic word embeddings for each temporal instance of the corpus. These embeddings were aligned with one another to ensure that terms from each temporal instance were comparable. The technique we used is based on TWEC [3]. It uses a negative sampling optimization of softmax to maximize the probability that a set of words surrounding word $w_k$ are representative of its context in time ($C^t$), when multiplied by the mean of atemporal word embedding vectors from $u$ (the compass) for the same set of context words around $w_k$.
\begin{displaymath}
  \mathnormal {\max_{ \mathbf{C}^t }\log P({w_k}|\gamma({w_k})) = \sigma (\vec{u}_k \cdot \vec{c} {\,^t_{\gamma(w_k)}})} 
\end{displaymath}

Since the TWEC embedding model did not account for phrases, we incoporated an additional step to indentify them. Phrases (including drug names) were then specially encoded to allow them to be treated like words. Figure \ref{fig:rankscore} shows TF/IDF verses the mean embedding distance to the compass for candidate therapeutics.

\begin{figure}[H]
  \includegraphics[width=0.49\textwidth]{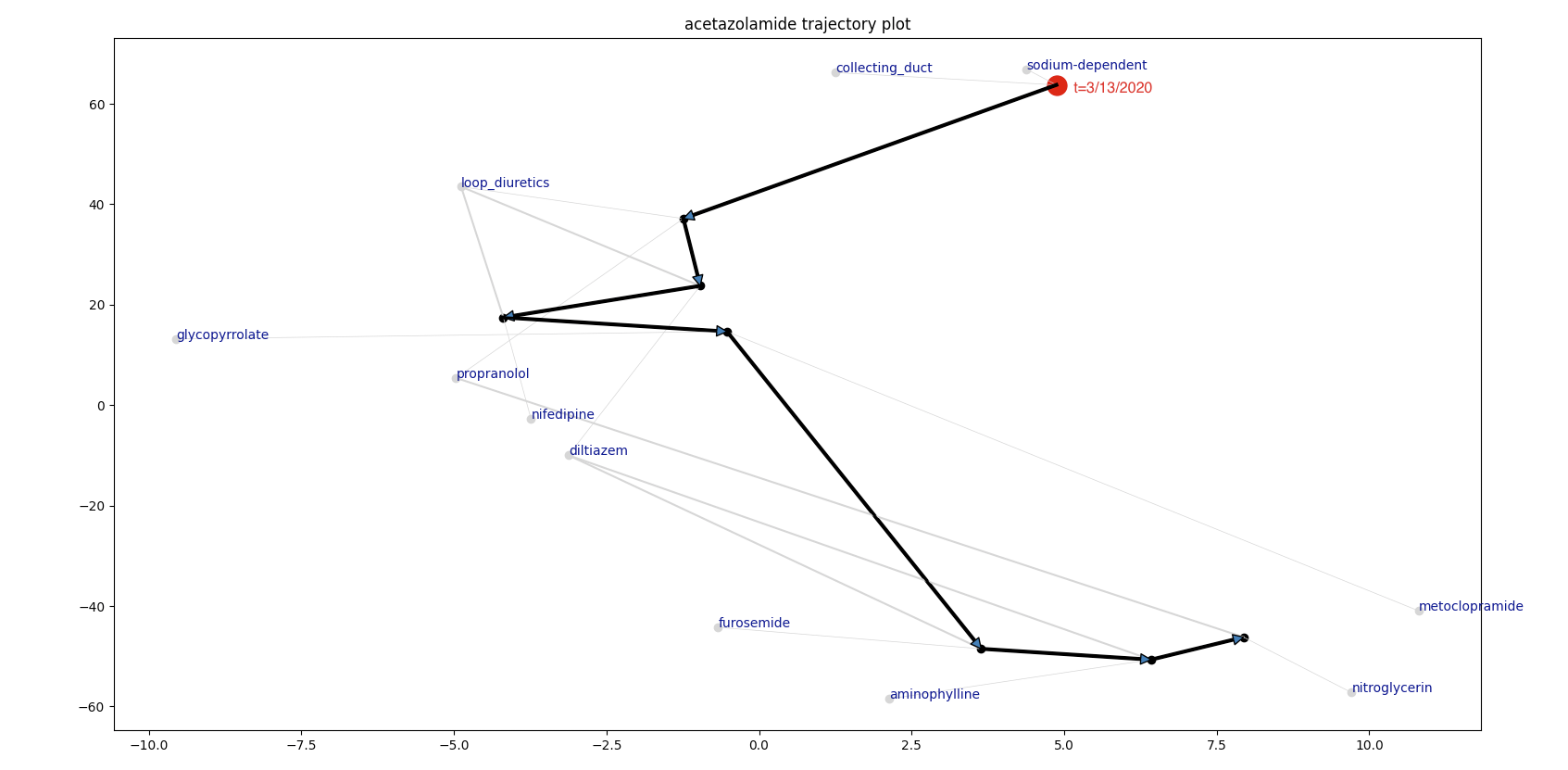}
  \caption{Semantic trajectory of 'acetazolamide' (rank -2.4). Nodes along the path represent the candidate embedding vector and its two closest terms at time $t$}
  \label{fig:trajectory}
  \Description{Semantic trajectory of 'acetazolamide' (rank -2.4). Nodes along the path represent the candidate embedding vector and its two closest terms at time $t$}
\end{figure}

Because diachronic embedding instances were aligned with one another, we were able to isolate a given candidate and visualize its semantic trajectory over time (Figure \ref{fig:trajectory}). As the trajectory is based on nearest neighbors at a given time, subtle changes in semantic associations become apparent \cite{stewart2017measuring}.

\section{Results and Analysis}

We detected 14\% (1267) of the candidate therapeutics in CORD-19 at 3/13, increasing to 26\% (2361) by 6/30. For candidates detected in multiple adjacent temporal instances of the corpus, we were able measure their changes over time. We found that many candidates exhibited increases in frequency, and stable or strengthening semantic associations. However, given the nature of this corpus, we suspected some would exhibit other kinds of change over time.

\begin{figure}[H]
  \includegraphics[width=.85\linewidth]{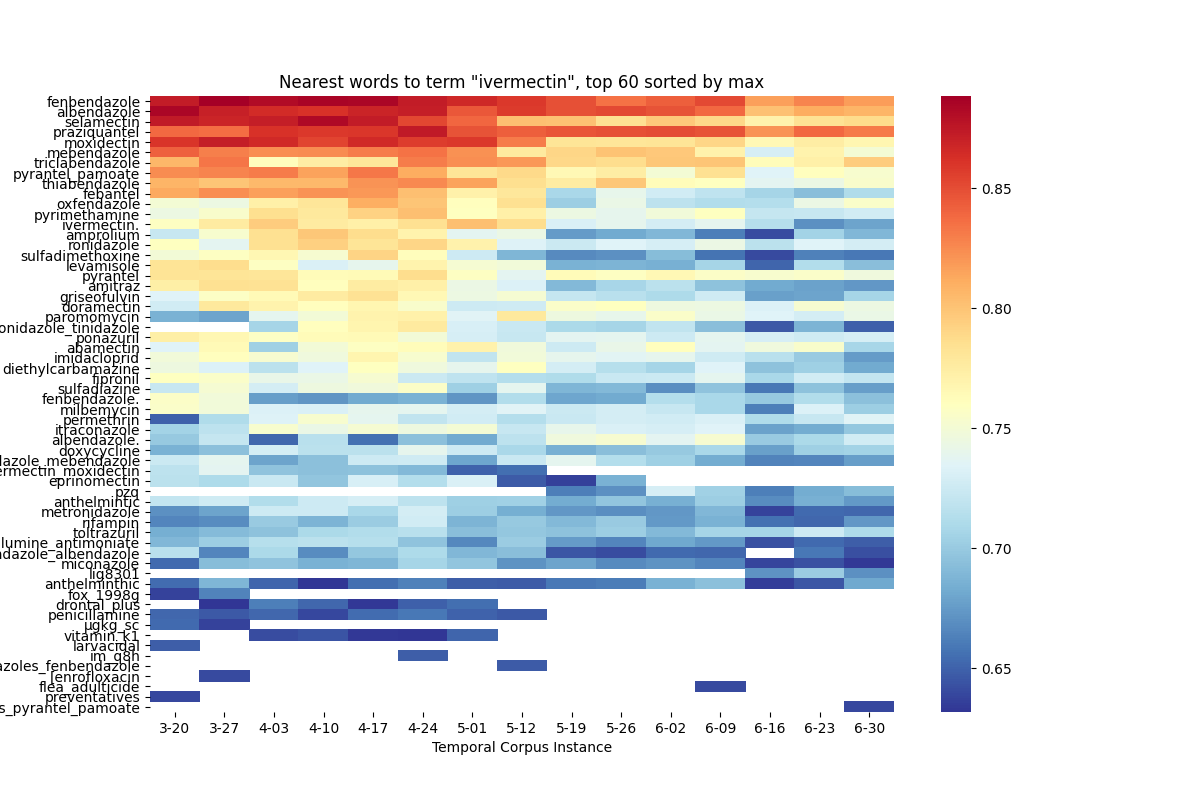}\\
  \caption{Example of weakening semantic associations for the candidate therapeutic \textit{ivermectin}}
  \label{fig:ivermectin}
\end{figure}

\begin{figure}[H]
  \includegraphics[width=.85\linewidth]{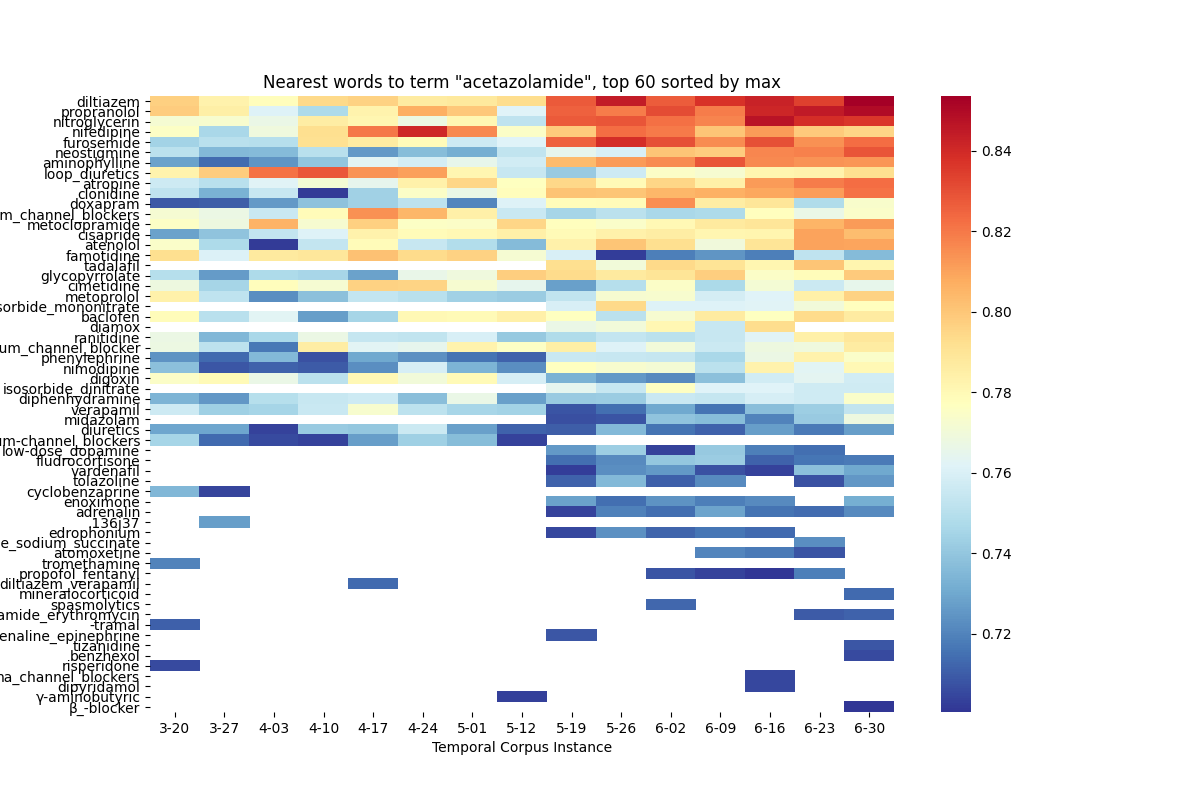}\\
  \caption{Example of disrupted semantic associations for the candidate therapeutic \textit{acetazolamide}}
  \label{fig:acetazolamide}
\end{figure}

We found that some candidate therapeutics exhibited different patterns of semantic associations. Using heatmap visualizations as described in \cite{Xu17}, we can illustrate two additional recurring patterns of behavior. Some candidates exhibited weakening semantic associations over time (Figure \ref{fig:ivermectin}), while others exhibited an abrupt persistent shift to a different pattern (Figure \ref{fig:acetazolamide}). Additionally, we found that these changes were not strongly correlated with changes to a target's frequency scores.

\section{Conclusion}
Our diachronic survey of candidate therapeutics for COVID-19 in the CORD-19 corpus found that some exhibited weakening or abrupt changes of semantic associations. We speculate that this could be related to the publication of new research that positively or negatively affected consideration of a candidate therapeutic as a treatment for COVID-19. Future work will investigate how to detect and quantify these patterns, and to determine if there are  any correlations between a target's rank and magnitude of change.

\bibliographystyle{ACM-Reference-Format}
\bibliography{covsensys}

\end{document}